# SMALL-SCALE MAGNETIC ISLANDS IN THE SOLAR WIND AND THEIR ROLE IN PARTICLE ACCELERATION.
# I. DYNAMICS OF MAGNETIC ISLANDS NEAR THE HELIOSPHERIC CURRENT SHEET

O. Khabarova[1], G. P. Zank[2,3], G. Li[2], J. A. le Roux[2,3], G. M. Webb[2], A. Dosch[2], and O. E. Malandraki[4]

1 Heliophysical Laboratory, Pushkov Institute of Terrestrial Magnetism, Ionosphere and Radiowave Propagation RAS (IZMIRAN), Troitsk, Moscow 142190, Russia
2 Center for Space Plasma and Aeronomic Research (CSPAR), University of Alabama in Huntsville, Huntsville, AL 35805, USA
3 Department of Space Science, University of Alabama in Huntsville, Huntsville, AL 35899, USA
4 IAASARS, National Observatory of Athens, GR-15236 Penteli, Greece

ABSTRACT

Increases of ion fluxes in the keV–MeV range are sometimes observed near the heliospheric current sheet (HCS) during periods when other sources are absent. These resemble solar energetic particle events, but the events are weaker and apparently local. Conventional explanations based on either shock acceleration of charged particles or particle acceleration due to magnetic reconnection at interplanetary current sheets (CSs) are not persuasive. We suggest instead that recurrent magnetic reconnection occurs at the HCS and smaller CSs in the solar wind, a consequence of which is particle energization by the dynamically evolving secondary CSs and magnetic islands. The effectiveness of the trapping and acceleration process associated with magnetic islands depends in part on the topology of the HCS. We show that the HCS possesses ripples superimposed on the large-scale flat or wavy structure. We conjecture that the ripples can efficiently confine plasma and provide tokamak-like conditions that are favorable for the appearance of small-scale magnetic islands that merge and/or contract. Particles trapped in the vicinity of merging islands and experiencing multiple small-scale reconnection events are accelerated by the induced electric field and experience first-order Fermi acceleration in contracting magnetic islands according to the transport theory of Zank et al. We present multi-spacecraft observations of magnetic island merging and particle energization in the absence of other sources, providing support for theory and simulations that show particle energization by reconnection related processes of magnetic island merging and contraction.

*Key words:* acceleration of particles – magnetic reconnection – solar wind – Sun: heliosphere –Sun: magnetic fields – turbulence

## 1. INTRODUCTION

The appearance of magnetic islands (or flux ropes) as a result of dynamical processes is an intrinsic property of plasma. Such structures with a rotating magnetic field and an anti- correlated "density–magnetic field" pair are commonly observed in laboratory plasmas as well as in the terrestrial magnetosphere (Moldwin & Hughes 1992; Waelbroeck 2009). The Sun is a natural source of large-scale magnetic islands as well. A Coronal Mass Ejection (CME) initially contains a huge flux rope (a magnetic cloud) easily observed with corona- graphs. As the Interplanetary Coronal Mass Ejection (ICME) propagates through the solar wind, it becomes more and more complicated in comparison with the original CME structure. A cascade of smaller-scale magnetic islands separated by current sheets (CS) is generated at the leading edge of the flow due to turbulence, reconnection, and numerous instabilities (Chian & Munoz 2011; Zharkova & Khabarova 2015).

The regular solar wind plasma is characterized, too, by the occurrence of magnetic islands of smaller sizes (in comparison with magnetic clouds; Bothmer & Schwenn 1998; Cartwright & Moldwin 2008, 2010). In general, magnetic islands originate from the dynamical processes of magnetic reconnection and turbulence (Greco et al. 2010; Markidis et al. 2013). There are two populations of solar wind magnetic islands with different distributions (Janvier et al. 2014). The first is associated with magnetic clouds, as discussed above. It consists of magnetic ropes with a typical size $l > 0.1$ AU and possesses a Gaussian-like distribution. On the other hand, a second small-scale ($l < 0.1$ AU) population, distinct from magnetic clouds, exhibits a power-law distribution.

This might indicate quite different origins for magnetic islands belonging to the different populations and confirms that a characteristic scale for these structures is a function of the ambient plasma conditions (Waelbroeck 2009).

Magnetic islands may possess various forms, from round to elongated, but significant elongation unavoidably leads to pinching that produces several structures with closed magnetic field lines separated by CSs. The largest such structures are called "flux ropes" or "magnetic clouds," but we are interested in the smallest-scale structures and we refer to them as magnetic islands to distinguish between structures of different origin that have different characteristic sizes. It is important to note that some investigators use the term "flux ropes" even for the smallest-scale formations of this sort.

The properties of magnetic islands observed in the solar wind such as their width and frequency of occurrence as well as their impact on the surrounding plasma depend on their location and origin. This is well known for magnetic clouds, but the properties of medium- (~0.1AU at 1 AU) and small-scale (~0.01AU at 1 AU) magnetic islands are poorly known.

Their role in particle dynamics is unclear, too, in part because of our insufficient understanding of the temporal and spatial evolution of these structures. The role of magnetic islands may be important for particle acceleration in the solar wind. Magnetic islands may experience merging or contraction, which leads to charged particle energization. Drake et al. (2006a), Oka et al. (2010), Bian & Kontar (2013), Zank et al. (2014), and le Roux et al. (2015) derived a transport equation for particles trapped in the vicinity of and within magnetic islands, experiencing repeated interactions with the electric field induced by the merging process and/or first-order Fermi energization via magnetic island contraction.

Local power-law solutions are predicted as a consequence of both the induced magnetic island electric field and island contraction (Zank et al. 2014), in accordance with numerous simulations of particle energization in a reconnecting plasma (Cargill et al. 2006; Drake et al. 2006a, 2006b; Pritchett 2006, 2007, 2008; Oka et al. 2010; Guo 2014). The power-law index for the accelerated particle distribution is predicted to depend on the Alfvén Mach number $M_a$ (the ratio of solar wind speed to Alfvén speed) and the ratio of the magnetic island contraction timescale to the particle diffusion timescale (Zank et al. 2014).

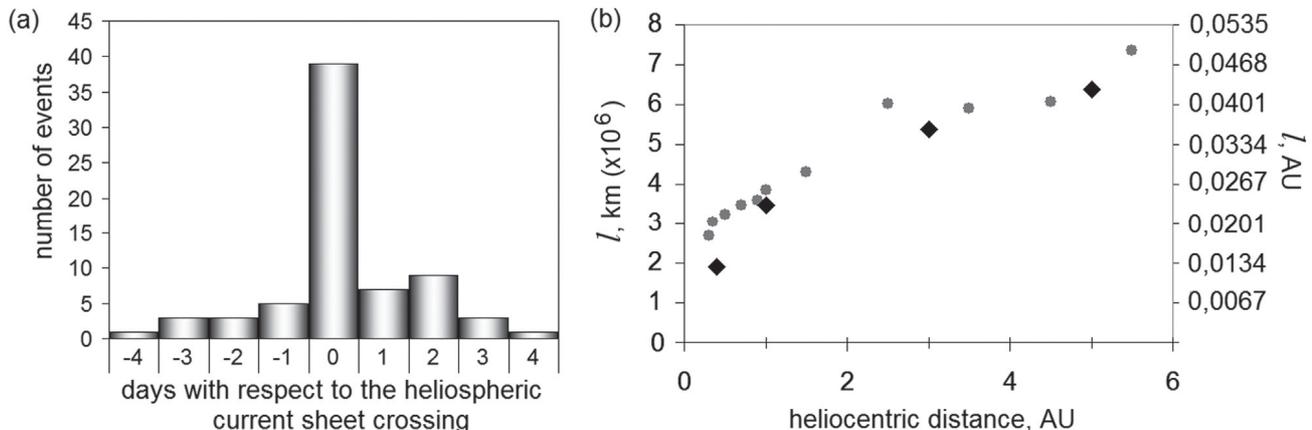

Figure 1. Spatial distribution of small-scale magnetic islands observed in the solar wind. (a) Frequency of observations of magnetic islands at 1 AU with respect to the heliospheric current sheet crossings, averaged over one day according to statistics by Cartwright & Moldwin (2010). 71 events were examined on the basis of WIND spacecraft measurements. (b) Averaged magnetic island width l vs. heliocentric distance as observed by Cartwright & Moldwin (2010; black diamonds) and calculated through the $l(B, r)$ dependence (1) (gray circles).

The Alfvén Mach number $M_a$ varies with the heliocentric distance as shown in Zank et al. (2014) and theoretically determines the power-law dependence on distance from the Sun. At the same time, $M_a$ experiences temporal changes at certain heliocentric distances. Finding direct evidence for the impact of magnetic island dynamics on particle energization throughout the heliosphere is a very difficult task, but it is possible to examine the plasma energization due to local magnetic island merging or contraction. To achieve this aim, we selected events in which particle acceleration was most probably determined by the occurrence of magnetic islands.

The place where $M_a$ significantly changes and magnetic islands are regularly observed is the heliospheric current sheet (HCS). It is easy to find that Ma is highly correlated with the plasma beta $\beta$ on all timescales. Strong variations of both parameters are observed around CSs; in particular, the correlation is best expressed around the HCS and may last for many hours.

The HCS represents a large-scale extension of the solar magnetic equator. Crossings of the HCS may be inferred from crossings of the neutral line of the interplanetary magnetic field (IMF), when the IMF (*B*) shows a clear sharp drop (because at least one of its components is zero). At the same time, the solar wind density *n* and plasma *β* increase, and the speed decreases slightly when high-resolution data is analyzed. On a large-scale (hourly or daily) perspective, this occurs during periods when the long-term background density and the IMF is enhanced (Khabarova & Zastenker 2011). The HCS identification can sometimes be complicated, mostly due to its irregular flapping and waving motion as well as its multifractal character. Since a spacecraft's velocity is much less than that of the solar wind, the HCS motion can manifest itself as multiple crossings by the spacecraft. The entire interval containing multiple crossings is often called the heliospheric plasma sheet.

The HCS is also referred to as sector boundaries in the sense that the magnetic field reverses across these structures. Investigators are usually interested in the approximate region where the IMF changes sign, simply identifying it as the HCS or sector boundary. Approximate times of sector boundary crossings at the Earth's orbit can be found in the list of sector boundaries crossings maintained by Dr. Leif Svalgaard: http://www.leif.org/research/sblist.txt .

Observations indicate that the HCS is surrounded by numerous secondary small-scale CSs separated by magnetic islands, most likely formed by continual reconnection along the HCS plane (Eastwood et al. 2002; Eriksson et al. 2014). Theoretically, the generation of a chain of magnetic islands via reconnection requires the presence of a guide field, consistent with observations (Markidis et al. 2013), which is naturally the case at the HCS.

A similar perspective is expressed by Bemporad (2008) and Susino et al. (2013), who argued that the very high temperature ions (Fe17+) observed in UV emission in a post-CME CS originate locally as a consequence of magnetic reconnection occurring through the entire CS. The authors believe that ongoing reconnection produces numerous magnetic islands around CSs in the corona and continues further in the solar wind. The islands near the CSs broaden with distance from the Sun, increasing the observed multifractality of the solar wind CSs, which may explain some features of time–intensity
profiles of post-ICME energetic particle events.

Not surprisingly then, small-scale magnetic islands are mainly localized around the HCS. Figure 1(a) is adapted from Figure 9 of Cartwright & Moldwin (2010) and shows the separation time between the HCS and the center of small-scale magnetic islands. According to Figure 1(a), the probability of finding small-scale magnetic islands increases with decreasing distance to the HCS. Another important result obtained by Cartwright & Moldwin (2010) is the dependence of island width on heliocentric distance r (shown as black diamonds in Figure 1(b)).

A magnetic island width *l* may be expressed as the empirical scaling

$$l \sim W^{1/2} \sim r(B)^{1/2} \quad , \tag{1}$$

where W is the magnetic flux. This formula is used frequently for island width estimates in tokamaks and stellarators (Waelbroeck & Fitzpatrick 1997; Fitzpatrick 1999; Waelbroeck 2009), indicating a possible commonality of processes that occur in different plasmas. By using an averaged *B* throughout the inner heliosphere as a function of heliocentric distance *r* (Khabarova & Obridko 2012), we can compute *l* (*B, r* ) and combine it with the Cartwright & Moldwin (2010) database (shown as diamonds in Figure 1(b)). Pioneer Venus Orbiter, Helios 2, IMP8, and Voyager 1 data were used for calculation of the *B(r)* dependence (see details in Khabarova & Obridko 2012). Our calculations yield the gray circles in Figure 1(b) and illustrate that the trend in size of magnetic islands with heliocentric distance is determined by the decrease in *B*.

In Figure 1(b), the width l is shown in both km and AU for the reader's convenience. To obtain *l* in kilometers, a multiplier of 1 AU (km)/100 should be applied to (1) if *B* is in nT and *r* is in AU. The empirical formula may, therefore, be represented as:

$$l = r(B/B_0)^{1/2} \quad \text{(km)}, \tag{2}$$

where $B_0 = 10\mu T = 10{,}000$ nT is the minimum coronal magnetic field magnitude, which is in accordance with estimates provided by Jensen & Russell (2009).

As discussed above, HCS crossings are associated with multiple secondary CSs and sharp changes of various plasma properties. When multiple CSs are present, the plasma becomes more intermittent. It has been shown that the plasma surrounding the HCS exhibits signatures of intermittent turbulence (Osman et al. 2014), and the magnetic field power spectra tends to become Kolmogorov-like when in a CS- abundant plasma (Li et al. 2011).

These features may be related to the presence of magnetic islands near the HCS and the magnetic reconnection process, but a direct correspondence has not yet been revealed. Particle acceleration in the HCS region has not been thoroughly examined before, but it is known from Ulysses observations that the energetic ion flux intensity increased every time the spacecraft crossed the HCS (Sanderson 1997; Lanzerotti & Sanderson 2001; see also http://www.esa.int/esapub/sp/sp1211/hspp4.htm). A similar correlation between increased fluxes of ions and HCS crossings was observed by the Voyagers in the outer heliosphere (Richardson et al. 2006).

Enhancements of energetic particle fluxes associated with the HCS are usually explained in terms of the dominant paradigm: all solar wind particles possessing keV–MeV energies are accelerated either at the Sun or at the ICME bow shock/interplanetary shocks. They propagate along magnetic lines, and their properties are determined by the source and depend on the magnetic connection (Li et al. 2003; Chollet & Giacalone 2008; Chollet et al. 2010). It is believed that ICME shocks begin to accelerate particles rather close to the Sun, which leads to uniform time–intensity profiles as observed at 1 AU (Zank et al. 2000, 2004, 2007; Reames 2009, 2013).

Even in the case when the source of acceleration is interplanetary shocks of another origin (for example, a coronal hole-related corotating interaction region (CIR)), accelerating structures are so wide that solar wind particles should experience strong scattering. As a result, at 1 AU, observers will detect approximately the same time–intensity profile of a solar energetic particle (SEP) event within ~30° angle. If two closely located spacecraft observe very different suprathermal particle characteristics, this is explained by different connections along magnetic field lines back to the Sun/the ICME (see, for example Chollet & Giacalone 2008; Chollet et al. 2010 and references therein). Unfortunately, such explanations are not always convincing, especially in cases when the separation angle between two spacecraft is very small. Tracing particles back to the source is a difficult task. Corresponding simulations impose numerous assumptions, and the results are never exact (because there is no possibility to check them experimentally). Therefore, the problem of "local SEP events" is still unsolved.

In summary, the most widely accepted explanation for the presence of energetic particles near the HCS is that it results from SEP propagation along the IMF from a source at the Sun or from a shock front. However, another possible source of accelerated particles is magnetic reconnection. Particle energization up to MeV energies due to magnetic reconnection has been experimentally observed and theoreti- cally explained for the case of explosive spontaneous reconnection in the Earth's magnetotail (Zelenyi et al. 1990; Ashour-Abdalla et al. 2011). However, this remains a matter of dispute because of a lack of observations.

Furthermore, Gosling et al. (2005a, 2005b) claimed the absence of energetic particle effects associated with magnetic reconnection exhausts in the solar wind, i.e., one can hardly expect to find significant local acceleration of particles at the reconnecting HCS. On the other hand, there is evidence that particles with energies of keV–MeV are associated with small-scale magnetic islands observed close to the HCS (Murphy et al. 1993). Local SEP events presented by Murphy et al. (1993) were short (lasting for several minutes) and weak (there was an energy cutoff at ~150 keV). They coincided with rotation of magnetic field in magnetic islands and followed heat flux dropouts, which is a possible signature of magnetic reconnection. Murphy et al. (1993) believed that the observations supported the theoretical results of Goldstein et al. (1986), who claimed that the maximum energy achievable during magnetic reconnection at the HCS should be ~100 keV.

Recently, Zharkova & Khabarova (2012) showed with particle-in-cell simulations that local particle acceleration due to a reconnection electric field in the presence of a strong guide field can explain a few key observational features in the supersonic solar wind, such as spatial profiles of ion velocities guided by the polarization (Hall) electric field induced by separated particles of the opposite charges and the behavior of suprathermal electrons around the HCS. The accelerated electrons can gain energies up to several keV, and ions – up to hundreds keV. The energies and pitch-angle distributions are found to be strongly dependent on the local configuration of the IMF near the region of the HCS that was undergoing magnetic reconnection. It was shown that some small- and medium-scale features of the IMF and plasma characteristics observed near the HCS may be explained simultaneously only if magnetic reconnection was assumed to occur recurrently in many places on the CSs (Zharkova & Khabarova 2012), which is entirely possible at the HCS.

The simulation of Zharkova & Khabarova (2012) applies to a region about 20 Larmor radii around the HCS, but accelerated particles can be observed much farther away. It is possible to explain this if one assumes that particles, initially accelerated via a magnetic reconnection process, are trapped in a region nearby the HCS, where they experience prolonged interactions and re-acceleration, achieving high energies. Taking into account the above results on the presence of magnetic islands in the vicinity of the HCS, we can suppose that magnetic islands provide the pool of accelerated particles and play a significant role in particle energization according to mechanisms discussed in Zank et al. (2014).

Given the emergence of theoretical models describing particle acceleration by reconnecting magnetic island processes (Drake et al. 2006a, 2006b, 2013; le Roux et al. 2015; Zank et al. 2014; Matthaeus et al. 1984; Hoshino 2012), numerous simulations (Wei et al. 2000; Cargill et al. 2006; Oka et al. 2010; Zharkova & Khabarova 2012, etc.), and indirect observations, it is becoming increasingly important to identify concrete observational evidence that particle acceleration is associated with small-scale magnetic islands in the supersonic solar wind.

Furthermore, it is important to address the apparent contradicting evidence related to particle acceleration in the vicinity of the HCS. In this paper, we identify specific cases of magnetic island dynamical evolution, including merging. We explore in detail the energetic particle characteristics in the vicinity of dynamically evolving magnetic islands, showing that a clear correspondence between energetic particles and magnetic islands indeed exists. This resolves, we believe, some apparently contradictory observations that show particle acceleration is sometimes and is sometimes not associated with regions in which reconnection is occurring.

2. OBSERVATIONS OF MAGNETIC ISLANDS AND PARTICLE ACCELERATION
2.1. Particle Acceleration Associated with Merging Magnetic Islands near the Reconnecting HCS

Figure 1(b) describes the typical island width versus heliocentric distance from the Sun. It is obvious that the islands experience dynamical changes at smaller scales. This may be the appearance and disappearance, merging and contracting of magnetic islands, possibly in response to local plasma conditions (Feng et al. 2012). We first show that the process of small-scale magnetic island merging occurs within a relatively short time period near 1 AU in support of the idea for additional sources of particle acceleration near the HCS (Zank et al. 2014; le Roux et al. 2015). We believe that magnetic islands located near CSs in the solar wind are most likely a consequence of magnetic reconnection occurring from the corona to several AU (Susino et al. 2013; Eriksson et al. 2014).

We analyze an event observed during a complicated (but rather typical) HCS crossing where multiple secondary CSs exist (Figures 2, 3). According to suprathermal electron pitch-angle histograms (see SWEPAM data http://www.srl.caltech.edu/ACE/ASC/DATA/level3/swepam/index.html), a change of direction in electron motion occurred on 1999 September 6, but the main crossing of the HCS was detected on 1999 September 7 as shown in Figures 2 and 3 (see also Leif Svalgaard's list). As mentioned above, the identification of the HCS from magnetic data is not an easy task because of the HCS' multifractal nature and its waving and flapping motion. It cannot be detected easily using the thin CS identification technique proposed, for example, by Li (2008) either, because it is impossible to distinguish between an ordinary CS and the HCS using the automatic CSs detection method.

The best way to identify the HCS remains visual inspection, which is unavoidable even if a variance analysis is employed (Sonnerup & Cahill 1968). An example of a typical crossing of the HCS as well as the way to detect it visually can be found in Zharkova & Khabarova (2012). In brief, the HCS possesses the same properties as any CS, which are a sharp decrease in the IMF strength (because at least one of the IMF components becomes zero), a high plasma beta, and increased density. Additional criteria should be applied, such as a sharp and relatively stable changing of the IMF azimuthal angle ($\phi$- angle), the existence of numerous secondary CSs nearby, and a sharp change in the suprathermal electron direction of motion, which is not necessarily a rotation by 180° because the main rotation occurs at the edge of the area filled with magnetic islands surrounding the HCS. This area can be considered part of the plasma sheet as well.

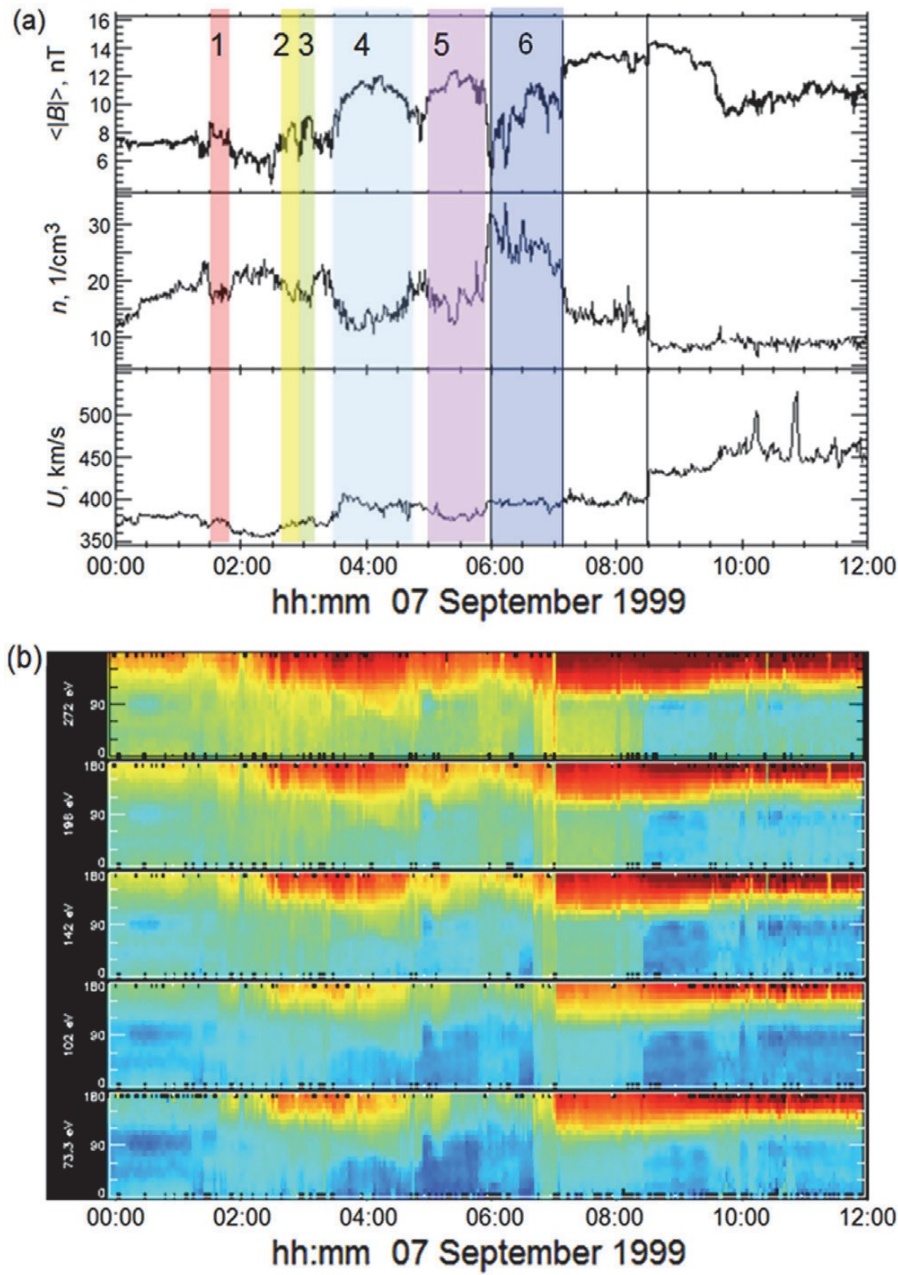

Figure 2. Merging magnetic islands observed by ACE in the vicinity of the heliospheric current sheet during the crossing of 1999 September 7: plasma and magnetic field parameters and suprathermal electron pitch-angle spectrograms.
(a) From top to bottom: the interplanetary magnetic field strength |B|, the plasma density $n$, and the solar wind speed $U$. Three very small-scale and three small magnetic islands are identified by colored strips. The vertical black lines identify the main crossings of the heliospheric current sheet. (b) ACE SWEPAM suprathermal electron pitch-angle distribution functions at different energies corresponding to (a). From top to bottom: 272, 196, 142, 102, and 75.3 eV.

The way to identify small-scale magnetic islands was proposed by Moldwin et al. (1995, 2000). This demands that bipolar field rotations coincide with a core field enhancement (an example of visual detection of these structures can be seen in Cartwright & Moldwin 2010). A chain of magnetic islands that is usually observed near the HCS may be simply characterized by anti-correlation of the solar wind density and the IMF strength as well as a visible rotation of the magnetic field vector, which can be seen from the IMF hodograms (see below).

The puzzling observation that electron pitch-angle spectrogram changes do not correspond to HCS signatures in IMF data has been discussed many times (see, for example, Kahler & Lin 1994, 1995; Crooker et al. 2004). It is thought that large-scale structures with a closed magnetic field rooted in the Sun (such as coronal streamer magnetic loops) may lead to a very complicated topology and be responsible for the observed mismatches.

We will discuss a set of events from an alternative perspective: the regular and frequent occurrence of "sector boundaries without field reversals and field reversals without sector boundaries" at 1 AU (Crooker et al. 2004, p. 1) means that this is a common feature of the HCS and certainly not exceptional. We suggest here an HCS structure that possesses small-scale ripples and is surrounded by magnetic islands. Secondary CSs are formed as boundaries of magnetic islands that are presumably produced by magnetic reconnection at the HCS.

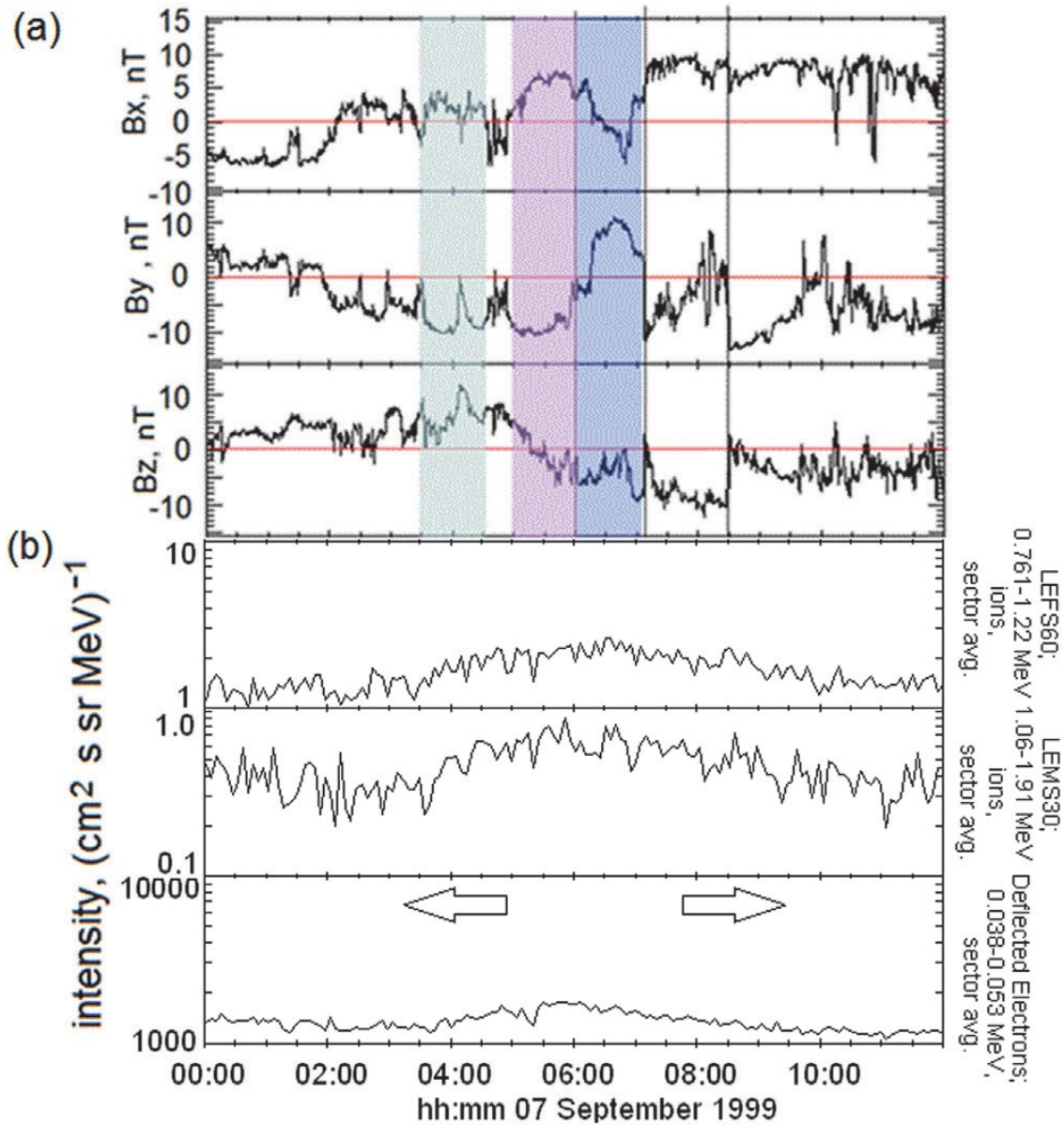

Figure 3. Merging magnetic islands observed by ACE in the vicinity of the heliospheric current sheet during the crossing on 1999 September 7: the IMF component variations in comparison with energetic particle flux increases. The three components of the interplanetary magnetic field in the GSE coordinate system for the period shown in the left panel. EPAM measurements of energetic particles are displayed in (b). The energetic particle flux increases are correlated with the occurrence of the largest magnetic islands.

The process of magnetic island formation and evolution during magnetic reconnection at CSs was investigated in many works (see, for example, Drake et al. 2006a, 2006b, 2013; Shibata & Tanuma 2001; Loureiro et al. 2007; Guo 2014). The resultant magnetic islands evolve and become detached from the "mother" CS, both drifting apart and making the main CS multifractal. The "mother" CS remains rather thick, but small-scale CSs separating the magnetic islands are much thinner. As a result, the reconnecting HCS is surrounded by a cloud of magnetic islands having their own small-scale magnetic separators. This leads to a situation when most observed changes in electron pitch angles are observed at the edge of a region filled with magnetic islands, i.e., rather far from the main crossing of the HCS. At the same time, electrons may experience scattering in this area, which is characterized by prolonged heat flux dropouts seen in suprathermal electron pitch-angle spectrograms.

In Figure 2, the IMF strength, the solar wind density, and speed (Figure 2(a)) are shown together with suprathermal electron pitch-angle spectrograms (Figure 2(b)). The behavior of the IMF components is compared with data from the Low Energy Magnetic Spectrometer (LEMS30) and the Deflected Electrons detector (DE) on board the ACE spacecraft in Figure 3.

As illustrated in Figures 2 and 3 magnetic islands are separated by small-scale CSs and may be identified easily by an increase of the IMF with an accompanying decrease in the solar wind density n or the dynamic pressure. Figures 2 and 3 show the HCS crossing on 1999 September 7 as observed by the ACE spacecraft with 16 s resolution for the magnetic field and 64 s resolution for the plasma. Figure 2(a) illustrates the possibility of multiple crossings of CSs interspersed with magnetic islands. The time of the HCS crossing is identified by the three vertical lines in Figures 2(a) and (b), suggesting that the HCS probably crossed the spacecraft position three times from 05:30 to 11:30 UT. This particular crossing was characterized by a highly dynamical HCS and multiple secondary CSs can be identified. The spacecraft detected at least nine strong and many small-scale CSs, which are boundaries of magnetic islands.

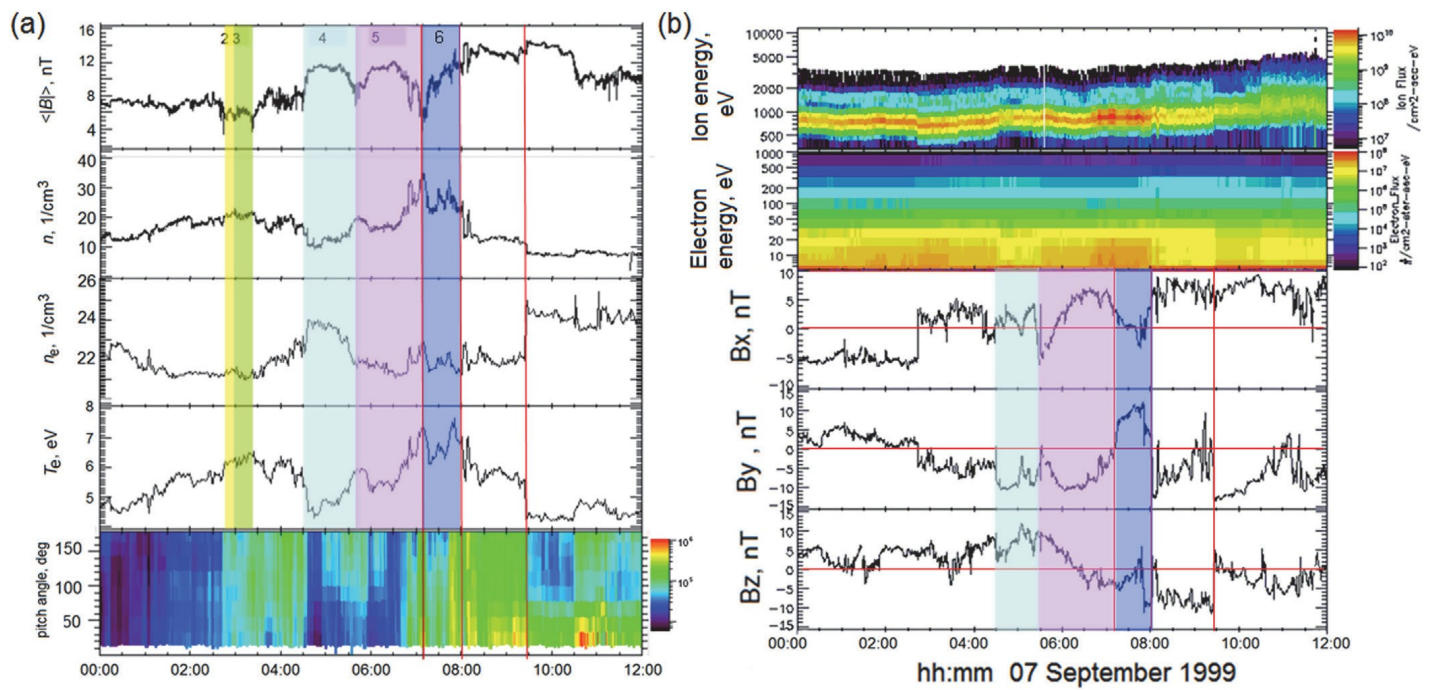

Figure 4. WIND observations of plasma and magnetic field variations across the HCS, previously seen by ACE (1999 September 7). The spacecraft were ~150 Re apart. Island merging with distance is seen. Suprathermal electron and ion fluxes vary in accordance with the occurrence of magnetic islands. The most intense fluxes are observed in magnetic island 6, which is bounded by two strong current sheets due to a ripple-like structure of the HCS.

The first very small-scale magnetic island is marked by the light-red strip following the crossing of the HCS-associated small-scale CS at 01:28:00 (Figure 2(a)). The crossing lasted approximately 25 minutes, which corresponds to a structure width of $5.6 \times 10^5$ km. Similar very small-scale magnetic islands, numbered 2 and 3 (marked with yellow and green in Figure 2(a)), were observed after crossings of two other secondary CSs at 02:32:30 UT and 02:54:30 UT, respectively. Two small-scale magnetic islands, separated by a strong CS detected at 04:52:09 UT, were observed at 03:32:57 UT and at 05:01:45 UT (the events numbered 4 and 5 respectively, marked by blue and purple strips). The magnetic island number 6 marked by a dark blue strip occurred between two possible HCS crossings. The width of the structures numbered 4–6 was ~$10^6$ km.

Each crossing of a strong CS corresponds to a variation of intensity in the spectrograms of the pitch-angle distribution of suprathermal electrons (Figure 2(b)). These changes manifest themselves as vertical lines because of the change in the direction of electron motion. The blue color used in Figure 2(b) indicates low intensity and the red color corresponds to high intensity. The color scale varies for each energy panel.

A detailed description can be found at http://www.srl.caltech.edu/ACE/ASC/DATA/level3/swepam/index.html. In this particular case, electrons of different energies exhibited variability in pitch angle before the main crossing of the HCS, which was observed at 07:06:00 UT (the second vertical line in Figures 2(a) and (b)). Such changes in the behavior of suprathermal electrons may be a sign of magnetic reconnection as discussed by Zharkova & Khabarova (2012). The small-scale magnetic islands numbered 4, 5, and 6 can be identified in the spectrograms as extended blue areas in the 75.3 and 102 eV energetic channels, indicating an absence of electrons with pitch angles <90°. A gradual increase of the suprathermal particle energy in the low-energy range (up to several keV for ions and 100 eV for electrons) with a maximum occurring exactly at the main HCS crossing is typically observed during HCS crossings at 1 AU (see Figure 5 in Zharkova & Khabarova 2012).

During the event shown in Figure 2, besides an energy enhancement in the low-energy range, some increase in the suprathermal particle energy was observed in the range up to 1.91 MeV (LEMS30) and 1.22 MeV (LEFS60) for ions and 0.38–0.53 MeV for electrons over a wide area near the HCS including the magnetic islands marked in Figure 3(a) and Figure 3(b). The arrows in Figure 3(b) show a period of additional energization of ions and electrons in an extended region around the HCS, including the magnetic islands that are identified in Figure 2. It is possible to trace temporal and spatial changes in the morphology of the magnetic islands 4, 5, and 6 during their propagation because the WIND spacecraft was ~150 Re from ACE along the X (GSE) direction (Figure 4). The HCS crossings, identified by vertical lines in Figures 2 and 3, were detected clearly by WIND.

Comparing Figure 4 with Figures 2 and 3, one can see the effect of an ongoing merging process between islands 4 and 5 on these structures. WIND observed the IMF with a 3 s resolution and the plasma with a 24 s resolution, and detected only small-scale islands convincingly. The very small structures (numbers 2 and 3) merged or (number 1) contracted. As seen from Figure 4(a), island number 3 swallowed island 2. Magnetic islands 4, 5, and 6 can still be observed in the suprathermal electron pitch-angle spectrogram (bottom of Figure 4(a)) and suprathermal ion and electron fluxes (top of Figure 4(b)). The merging and elongation of the island structures resulted in a widening of magnetic islands 4 and 5 as well as the X-line area between them. Like the ACE measurements, the most intense fluxes of energetic particles were associated with island 6, confined between two HCS crossings.

In Figure 4(a) we also show WIND measurements of electron density and temperature. Variation in the electron density across the CS may provide additional evidence for magnetic reconnection. Usually, the electron and ion pressures are approximately equal, but magnetic reconnection demagnetizes the plasma and breaks the cylindrical symmetry of the electron pressure tensor. Such behavior is a consequence of a demagnetized electron diffusion region with increased electron temperature Te (Scudder et al. 2012). In an undisturbed plasma far from CSs, the electron and proton densities $n_e$ and $n$ vary simultaneously. During the event discussed here, $n_e$ and $n$ were anti-correlated up to the HCS crossing, and Te increased around the CSs (Figure 4(a)). A possible explanation of this effect is a separation of particles of opposite charges around the reconnecting CS (Zharkova & Khabarova 2012). Other evidence for magnetic reconnection is an increase in the solar wind speed, temperature, and density, corresponding to a reconnection exhaust. Sometimes a minimum variance analysis of the magnetic field is employed (Gosling et al. 2005b, 2007) to reveal the exhaust features, but in clear cases the exhausts can be seen simply through the mentioned parameter changes in ordinary coordinate systems not being oriented to the CS front (Enzl et al. 2014). Such an increase in the parameters obviously occurred, for example, during the third crossing of the HCS (see the behavior of corresponding parameters, except for the proton temperature which is not shown, at the moment of the third HCS crossing indicated with the vertical line around 8:30 for ACE and 9:30UT for WIND in Figures 2–4).

Evidence for the rotation of the IMF inside magnetic islands is illustrated in Figure 5. The hodograms, plotted for events 4, 5, and 6, clearly display the rotation of the magnetic field in the Bx–By plane (the ecliptic plane) and in the By–Bz plane.

In summary, ACE and WIND observations in the vicinity of the HCS both reveal the existence of magnetic islands of different sizes in the supersonic solar wind, as well as the merging of magnetic islands associated with increased energetic particle flux, especially pronounced in the case of plasma confinement between two strong CSs.

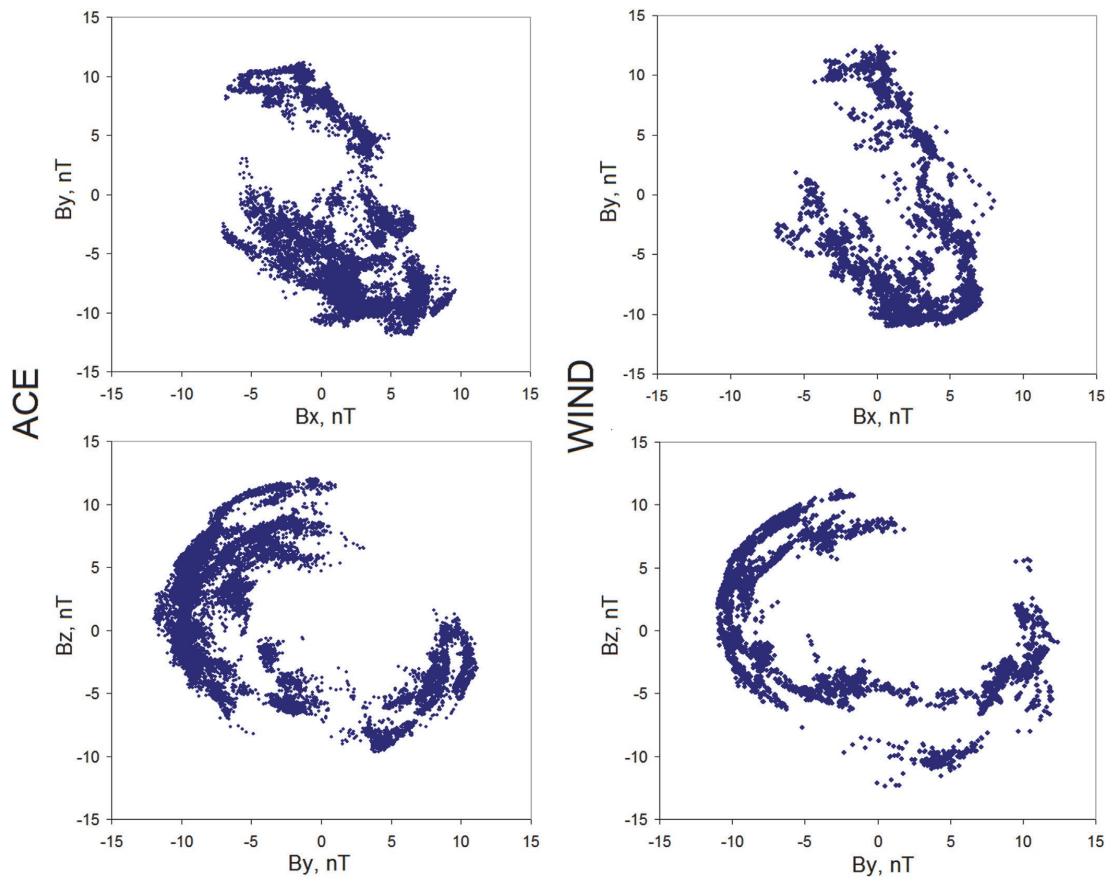

Figure 5. Hodograms of the interplanetary magnetic field for the time periods corresponding to observations of the magnetic islands 4, 5, and 6 (Figures 2 and 3) as seen by ACE (left) and WIND (right). Rotation of the magnetic field vector is clearly observed. Left panels: ACE measurements from 03:03:21 to 07:14:59 UT (1 s resolution). Right panels: WIND measurements from 04:20:01 to 08:01:58 (3 s resolution).

2.2. Particle Acceleration in Magnetic Islands Confined by an ICME and the HCS: Post-CME Case

The 2002 May 23 SEP event is unusual in exhibiting very high values of IMF and plasma parameters at 1 AU. This is a case where the CME trails an HCS and interacts with it in the way shown in the simulation by Manchester et al. (2014). They supposed that the magnetic flux rope inside an ICME can reconnect with the IMF not only at the leading edge (Chian & Munoz 2011), but also at the back side of the ICME. The HCS crossing occurred on 2002 May 19 (see L. Svalgaard's list), but a series of subsequent and complicated crossings of the HCS is found to continue up to the beginning of 2002 May 23, when a coronal hole stream reached the Earth. The corresponding drop in the solar wind density at about 4 am is seen in Figures 6(a) and (b), but the speed change is not so obvious because of the following shock and speed jump up to 1000 km s$^{-1}$ due to the ICME. As the ICME approached, precursor high energy particles (Figure 6(a)), locally accelerated by the shock (identified by vertical black line in Figures 6(a) and (b)), were observed. The energetic particle intensity then peaked at the shock, signaling that the shock is still accelerating particles to this energy at 1 AU (e.g., Malandraki et al. 2005). Such behavior of energetic particles is completely characteristic of diffusive shock acceleration at a CME-driven shock wave (Zank et al. 2000; Zank et al. 2007; Verkhoglyadova et al. 2010, 2012). This is typically seen in SEP events and ESP (energetic storm particle) events (Desai & Burgess 2008) and has been modeled by Li et al. (2003). Crossings of several CSs during the event are marked by vertical red lines.

Besides strong (but rather typical) particle acceleration at the shock, there was a second clearly identifiable SEP event within several hours of the shock passage and after the passage of the large-scale magnetic cloud. Its maximum is indicated with the purple stripe in Figures 6(a) and (b). This case was discussed by Tessein et al. (2013), who identified an unusual increase of energetic particle fluxes. They concluded that the increased energetic particle flux was probably due to a local acceleration process.

Two sharp drops of B are observed simultaneously with enhanced n on the edges of the purple stripe in Figure 6(b), corresponding to crossings of strong CSs. Comparing the IMF behavior with the plasma variations, we can conclude that it was a crossing of the HCS, now restored, after the ICME passage. As a result, the plasma was confined by the HCS from one side and by the ICME cloud from the other. This produced a very strong SEP event comparable with the ICME shock-related event. The rotation of the IMF vector in the corresponding area (indicated by purple stripe in Figure 6(c)) is shown in Figure 6(d). The very clearly rotating IMF vector is consistent with the existence of a strong magnetic island during the selected time period.

In summary, we conclude that the confinement of plasma that contains magnetic islands leads to strong particle acceleration in regions behind an enhanced magnetic field. Such events can provide insight into similar processes taking place closer to the Sun, where both the IMF strength and plasma density are much larger than at 1 AU.

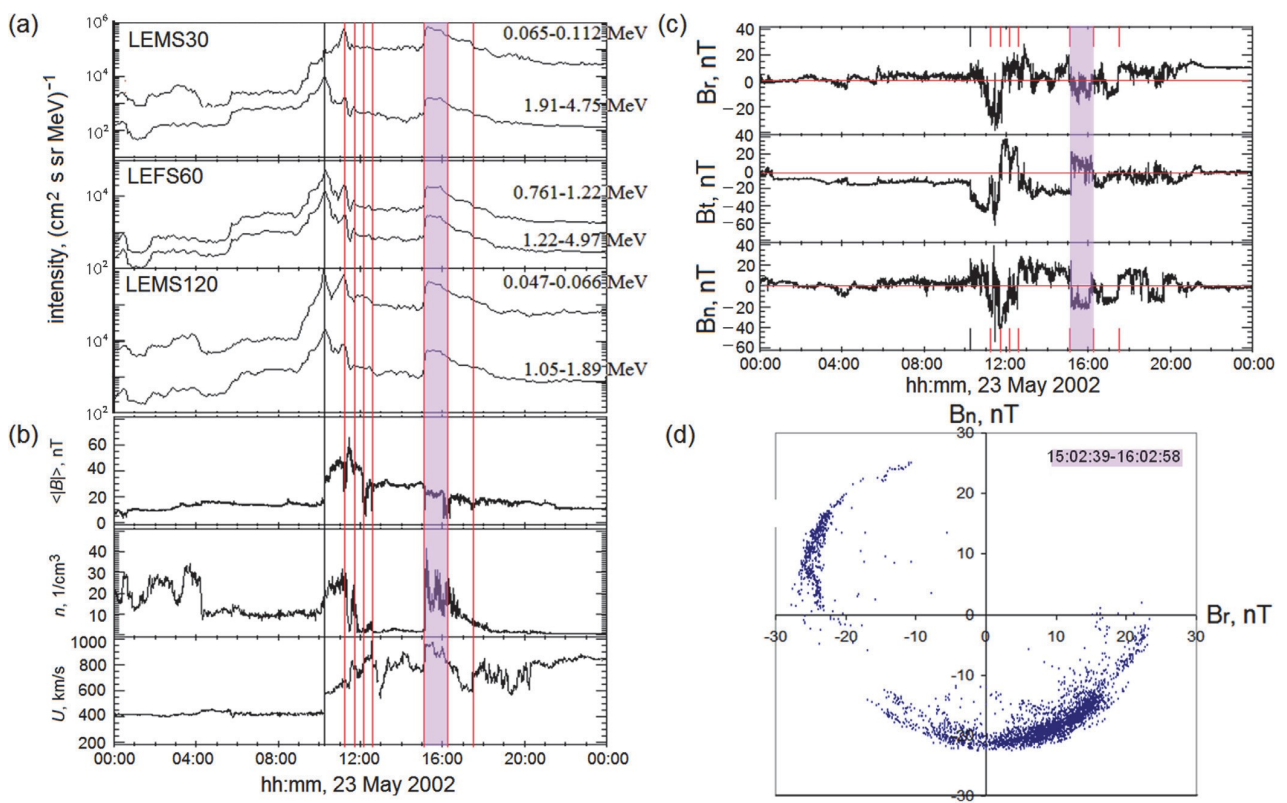

Figure 6. Unusual double SEP event observed by ACE on 2002 May 23. The first peak of energetic particle fluxes (a) corresponds to acceleration at the ICME shock. The second peak is related to the heliospheric current sheet crossing. (b) From top to bottom: the average strength of the IMF, the solar wind density, and speed. The black vertical line identifies the shock. Red vertical lines are current sheet crossings. The purple stripe indicates post-CME plasma confined by the HCS. (c) Three components of the IMF in the RTN coordinate system. Short lines correspond to the vertical lines in (a) and (b). (d) Hodogram of the magnetic field vector for the purple stripe period in (a), (b), and (c). IMF vector rotation perpendicular to the ecliptic plane is observed, which is a definite signature of a magnetic island.

2.3. Particle Acceleration in Magnetic Islands Confined by an ICME and the HCS: Pre-CME Case

The 2007 May 23 SEP event is famous for being observed by the very closely placed STEREO pair as well as by spacecraft at the L1 point. Despite such spacecraft configuration, they measured quite different time–intensity profiles of the energetic particle flux (shown in Figure 7, between events 1 and 2). A curiosity of this event was that accelerated particles sometimes behaved as though they originated from different sources, instead of simply along the same ICME shock front. Theoretically, the shock front should not change dramatically within the spacecraft separation angle of about 10° (just as in the case of a quite similar previous ICME that produced a classic SEP event just four days before on 2007 May 19).

The standard interpretation of the observed discrepancies is that (1) the only source of particles is that accelerated directly in the solar wind (by the CME-driven shock), and (2) that the ICME is magnetically connected to the Sun. The combination of these two paradigms gives only one possible explanation for the strange particle behavior, which is "the markedly different connection point of the field lines inside the ICME than outside" (Chollet et al. 2010, p. 9).

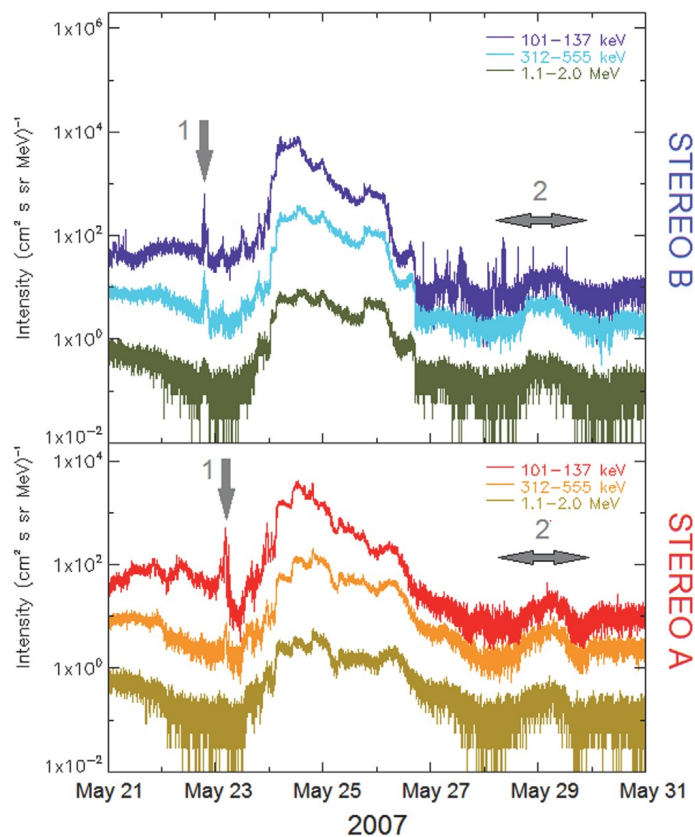

Figure 7. Event 1 shows an example of an unusual intense spike of strong particle energization near the HCS, which was deformed by the incident ICME. Event 2 corresponds to the case of small-scale magnetic island occurrence inside a confined formation not associated with the HCS (see the next article (part 2) in this series).

This scenario is of course possible, but the particle characteristics are not typical of particle transport with which we are familiar. Alternative explanations for the unusual energetic particle characteristics of the 2007 May 23 event are possible. We will not discuss the whole event here, but note that this particular ICME was preceded by the HCS. If we account for an increase of the reconnection rate at the HCS due to an external driver (for example, an ICME), the complicated ICME–HCS interaction, and the ability of the HCS to accelerate particles, it is clear that a very ordinary ICME may drive an extraordinary SEP event during specific conditions in the interplanetary medium (Chian & Munoz 2011; Zharkova & Khabarova 2015).

The intensity profiles of Figure 7 for STEREO A and B exhibit two strange and unexplained sharp increases. The duration of event 1 in Figure 7 was several hours. Event 1 was first observed by STEREO B and then by STEREO A (as seen in Figure 8). Perhaps, it was caused by the HCS ahead of the ICME. A similar but smaller increase in the intensity of energetic particles was observed after the ICME passage (event 2 in Figure 7). This was associated with the confinement of a large-scale area filled with magnetic islands. We will discuss this event in the second part of the present paper. When a low-energy solar wind structure is observed by two separated spacecraft with some delay, it is generally interpreted as being due to the propagation of the same structure to a different location.

Let us consider the variations of the suprathermal particle fluxes in detail. The HCS crossing was detected on 2007 May 22 at about 04:15 UT by STEREO B and at about 12:10 UT by STEREO A (see the left and right panels of Figure 8). There were several crossings of the HCS by the spacecraft during the next several hours, and both spacecraft observed two clear HCS crossings, evident from the back-ground of the enhanced IMF strength and density (especially for STEREO B), probably related to the approaching ICME (Figure 8).

STEREO B detected energetic particles several hours earlier than STEREO A (see the red vertical lines in Figure 8). The region indicated by the purple stripe in Figure 8 was bounded by two HCS crossings and characterized by the occurrence of multi-scale magnetic islands temporally coinciding with the increased energetic particle fluxes (see the two upper panels). It should be noted that the flux intensity coloring is different for STEREO B and STEREO A. The yellow color for STEREO B corresponds to the green for STEREO A in the upper spectrum figure (the Sunward ion flux) because STEREO B intensities were lower in that direction. At the same time the yellow color for STEREO B is red for STEREO A in the lower spectrum figure, as the fluxes of particles from the anti-Sunward direction were much more intense at STEREO B than at STEREO A. The strong anisotropy of the accelerated particles illustrates the dependence of particle energization on the local configuration of the magnetic field. In particular, plasma confinement by CSs and the increased surrounding density and the IMF seem to play a significant role, enhancing particle energization.

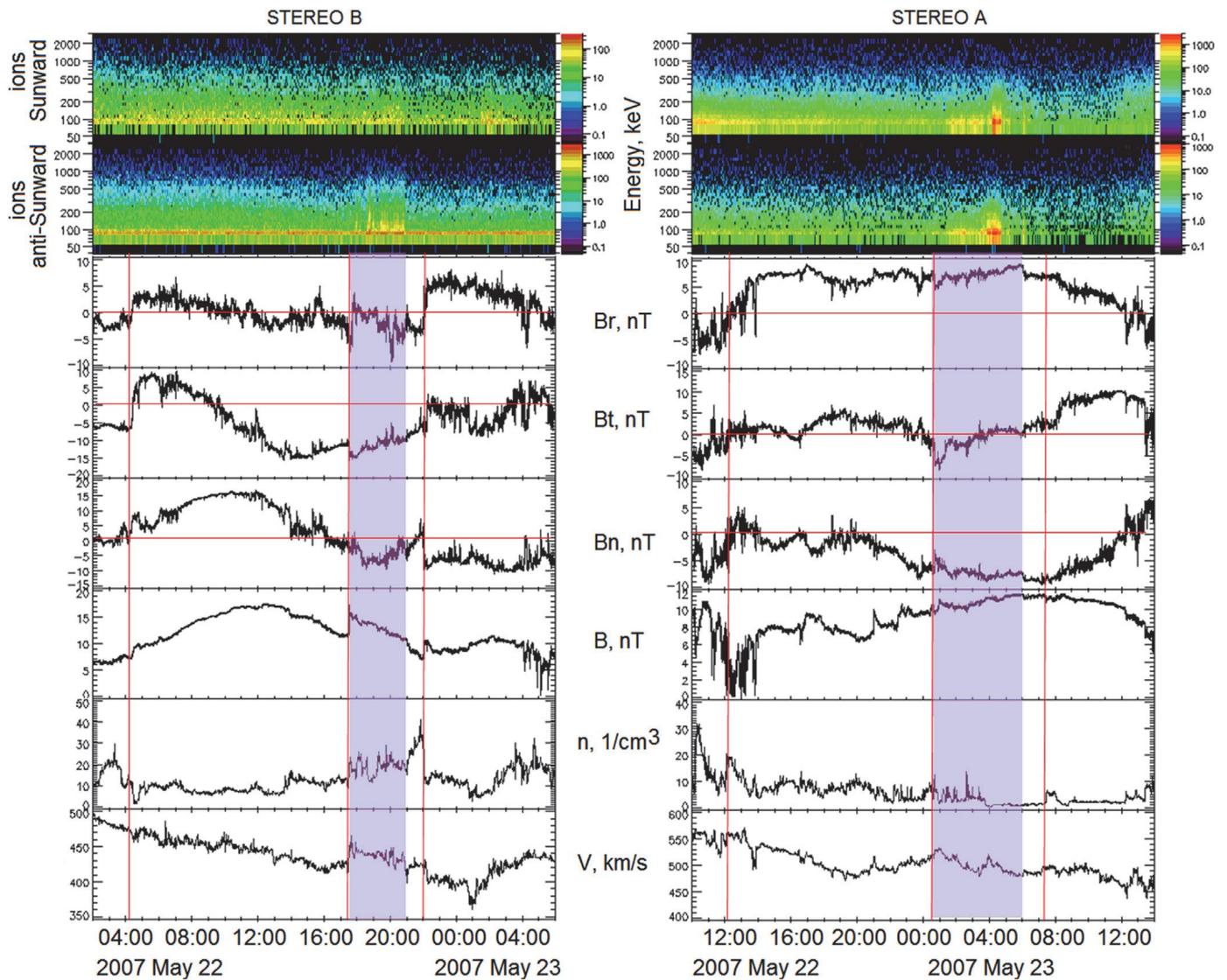

Figure 8. Example of local increase of energetic particle flux intensity (ions) observed by STEREO B (left) and STEREO A (right). Upper two panels: energy of Sunward and anti-Sunward ions (spectrogram). Then, from top to bottom: radial, tangential, and normal components of the IMF (in RTN coordinates; plasma density and speed). The HCS main crossings are indicated by vertical lines, and the flux enhancements are shown by purple vertical strips.

## 3. DISCUSSION

All the cases presented here reveal that approximately the same configuration is associated with an increased intensity of energetic particle fluxes. This is a relatively small-scale (about $10^6$–$10^7$ km) area filled with magnetic islands and bounded by two strong CSs. The existence of such an HCS structure is confirmed by ground-based interplanetary scintillation data from the Solar Terrestrial Environment Laboratory (STEL). The solar wind velocity profiles provided by STEL reflect the approximate IMF configuration as well (Jackson 2012, p. 69), revealing a large-scale waving of the HCS in which small-scale ripples exist.

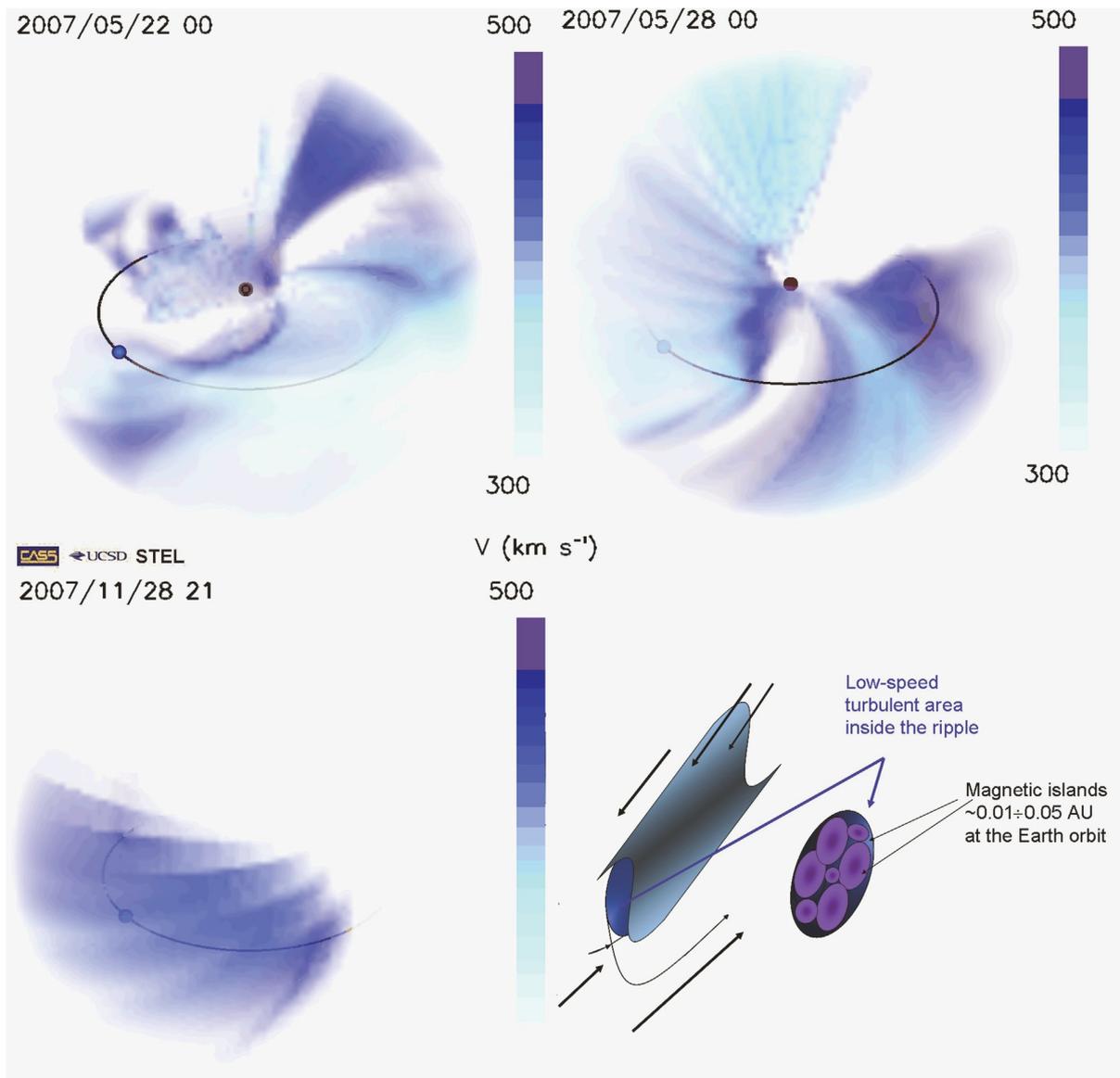

Figure 9. Heliospheric current sheet ripples. Blue figures—STEL observations drawn from interplanetary scintillation data. A cartoon shows a possible configuration of the magnetic field lines around the ripple as well as magnetic island confinement inside the ripple.

The upper panel of Figure 9 shows 3D STEL reconstructions for the period of the event displayed in Figures 6 and 7, and a clear example of a rippled configuration for the HCS is illustrated in the lower panel, together with a cartoon that indicates the place of small-scale magnetic islands. When such a structure moves through the spacecraft location, it resembles crossings of strong CSs, separated by magnetic islands.

The origin of a rippled HCS structure is still unclear. Perhaps, this is a consequence of some form of resonance in the rotating sheet or post-CME waving of the HCS due to reconnection, which is frequently observed both in magnetospheric and laboratory plasma (Frank & Satunin 2014; Le et al. 2014; Sitnov et al. 2014), but additionally complicated by rotation of a very large-scale CS in the case of the solar wind. As a whole, the STEL observations are in agreement with results obtained by Merkin et al. (2011), who showed the existence of the HCS ripples as a representation of a continuation of small-scale bending of the solar magnetic equator. A high-resolution simulation of the development of small-scale folds and ripples on the surface of the HCS with distance was performed, displaying a fine structure of the rippled HCS.

It is important to note that, according to simulations (Czechowski et al. 2010; Merkin et al. 2011), closed magnetic field structures such as loops or folds are relatively rare and unstable at 1 AU and further because of the radial expansion of the solar wind. Meantime, such structures are believed to be the main cause of the observed mismatches between polarity reversals in field data and in suprathermal electron pitch-angle spectrograms (see Section 2.1). It was found that two processes may be responsible for the HCS-fold disruption: (a) non-radial flows that push the HCS upward/downward and (b) magnetic reconnection.

The existence of the HCS ripples shown in Figure 9 based on 3D observations gives an easy explanation of the effect of too frequent multi-crossing of the HCS at 1 AU as well as the abovementioned mismatches. Furthermore, it does not require a hypothetical and unnatural pinching and bending of the HCS back to the Sun or the presence of structures with closed IMF to interpret spacecraft observations of the IMF and particle intensity profiles.

Depending on how a rippled HCS surrounded by magnetic islands is crossed, an observer will see different kinds of suprathermal electron pitch-angle spectrograms. For example, if a spacecraft crosses the HCS approximately perpendicularly to the HCS plane, sharp changes in suprathermal electron pitch angles at the edge of the area filled with islands and small-scale CSs will be observed. This will be followed by unstably directed suprathermal electron propagation characterized by heat flux dropouts and bi-directional strahls corresponding to crossings of magnetic islands separated by CSs. The main crossing of the HCS will be identified from magnetic field data and will be located somewhere in the middle of the area with the characteristics described above. Since the HCS front is usually inclined to the ecliptic plane, the HCS position may be observed as being not exactly in the middle of the plasma sheet.

Moreover, one can imagine the case when a spacecraft enters an area filled with magnetic islands from one side of the HCS, but does not cross the HCS. In this case the sector change will be seen in the suprathermal electron spectrograms, but the magnetic data will not confirm a sector boundary crossing.

Due to its rippled structure, the HCS is sometimes reminiscent not of a long-waving bashful ballerina's skirt (in Mursula and Hiltula's apt words; Mursula & Hiltula 2003), but rather a plissé skirt. The small-scale HCS ripples have a characteristic width of ~0.01–0.05 AU at 1 AU, and their size should grow with distance.

It is quite possible that the small-scale islands of maximum width observed by Cartwright & Moldwin (2010) reflect the case when a magnetic island fills the whole ripple as in Figure 5. It is interesting to note that the observed average scale size of small-scale magnetic islands at 1 AU (see Figure 1) agrees approximately with the magnetic turbulence correlation length (Matthaeus et al. 2005).

This agreement is consistent with previous work that concluded that low-frequency magnetic turbulence in the supersonic solar wind is dominated by quasi-2D turbulence (e.g., Matthaeus et al. 1990; Zank & Matthaeus 1992, 1993; Bieber et al. 1996). In addition, the dependence of the magnetic turbulence correlation length on radial distance determined from Voyager observations (Smith et al. 2001) agrees with the radial variation of the scale size of small-scale magnetic islands.

The rippled structure of the HCS effectively confines plasma and represents a natural tokamak or stellarator, which leads to local particle acceleration (Xu et al. 2002). Properties of magnetic islands in tokamaks have been studied for many years because they distort the plasma flow and convert toroidal into helical flux surfaces, which must be avoided (Waelbroeck & Fitzpatrick 1997; Fitzpatrick 1999; Fitzpatrick et al. 2006; Waelbroeck 2009; Fitzpatrick & Waelbroeck 2012). Our laboratory knowledge of magnetic island dynamics may be useful in analyzing dynamical processes in the solar wind plasma.

Particle acceleration due to magnetic reconnection at the HCS enhances the tokamak effect and allows particles to gain higher energies. The presence of turbulence and instabilities (Drake et al. 2006a, 2006b; Zharkova et al. 2011; Lazarian et al. 2012, 2015) as well as the formation of magnetic islands (Zank et al. 2014; le Roux et al. 2015) can significantly change the distributions of energetic particles and energies they gain in the vicinity of these CSs. Zharkova & Khabarova (2012) have demonstrated that a single flat HCS can accelerate particles to high energies due to magnetic reconnection. However, such cases of clear undisturbed CS in the heliosphere are rare. Most frequently, a wavy, rippled, and multifractal structure of the HCS surrounded by turbulent plasma is observed. In this case, mechanisms of particle acceleration related to dynamical magnetic islands become more appropriate (Zank et al. 2014; le Roux et al. 2015).

Moreover, for the case of a simple, flat HCS configuration, particle acceleration might not be particularly effective because of the absence of a large region filled with magnetic islands and plasma confinement enhancing particle acceleration. This produces a reasonable explanation for why an increase in energetic particle fluxes is sometimes observed, and sometimes not. The case of a very clear HCS crossing (as demonstrated in Gosling et al. 2005) is unlikely to yield significant particle acceleration near the HCS. In other words, one can expect to observe stronger acceleration of particles in the vicinity of rippled or disturbed HCS surrounded by numerous magnetic islands rather than in the case of a clear HCS crossing.

Particle acceleration due to the presence of magnetic islands near the HCS and smaller CSs may also be responsible for observations of energetic particles ~1MeV near CIRs or Stream Interaction Regions. The shock fronts of CIRs, commonly formed at heliocentric distances beyond 1 AU, are considered to be the main source of energetic particles at 1 AU in the absence of flares or ICMEs (Richardson 2004; Gómez-Herrero et al. 2011). The more intense and longer duration CIR MeV/nucleon ion enhancements have been typically associated with the reverse shock (e.g., Malandraki et al. 2008).

However, detailed analysis shows discrepancies and difficulties in interpreting certain events, because increases of energetic particle flux are usually observed in unexpected regions, for example, many hours before the CIR front arrival to a spacecraft. Furthermore, CIR-associated shocks are not as strong as CME-driven shocks. However, such events can be easily explained if we recognize that CIRs are often preceded by the HCS (Crooker et al. 1999). In this case, bounding the HCS-associated magnetic islands by the HCS on one side and a CS inside a leading CIR front on the other side may produce significant particle acceleration. We will discuss this idea and related results in the second part of this paper.

4. CONCLUSIONS

We have presented observations that show the occurrence of small-scale magnetic islands and related plasma energization in the vicinity of the HCS. We have found the following:
(1) magnetic islands in the solar wind possess a range of spatial scales;
(2) the characteristic width of magnetic islands varies with heliocentric distance as a square of the IMF strength;
(3) magnetic islands experience dynamical merging in the solar wind;
(4) increases of energetic particle fluxes in the keV–MeV range coincide with the presence of magnetic islands confined by strong CSs (the stronger the background magnetic field and density, the stronger the observed particle acceleration);
(5) the interaction of ICMEs with the HCS can lead to significant particle acceleration due to plasma confinement;
(6) the rippled structure of the HCS is confirmed on the basis of observations. Since such a structure confines plasma, it makes possible the strong energization of particles trapped inside small-scale magnetic islands.

All events that we examined were characterized by similar configurations of the magnetic field, i.e., the area filled with magnetic islands was bounded typically by two magnetic walls such as the HCS from one side and CSs at the leading (or trailing) edge of an ICME from the other side. Alternatively, the HCS itself can confine the magnetic islands because of its plissé-like profile.

The presence of magnetic islands in a wide area around the HCS can easily explain the changes observed in suprathermal electron pitch-angle spectrograms and sharp changes in azimuth IMF angle sometimes many hours before the HCS crossing, because electrons are scattered at magnetic islands. The most prominent change in electron pitch angles occurs at the edge of areas filled with magnetic islands.

In summary, initial particle acceleration due to magnetic reconnection at the HCS may be insufficient to obtain high energies, but the presence of magnetic islands inside the ripples of the HCS or between two CSs with a strong guide field offers the possibility of re-accelerating particles in the ways discussed in Zank et al. (2014) and le Roux et al. (2015).

We thank the SMEI's mission team for providing SMEI and STELab data on the official website http://smei.ucsd.edu/new_smei/data&images/data&images.html. The Solar Electron and Proton Telescope STEREO data were taken from the http://www2.physik.uni-kiel.de/STEREO/index.php (University of Kiel, Germany); other STEREO data were obtained from the STEREO science center official website http://stereo-ssc.nascom.nasa.gov/data.shtml . High-resolution ACE and WIND data as well as Pioneer Venus Orbiter, Helios 2, IMP8, Ulysses, and Voyager 1 data were taken from the official Goddard Spaceflight Center OMNIweb plus Web site: http://omniweb.gsfc. nasa.gov. We acknowledge the partial support of NASA grants NNX08 AJ33G, Subaward 37102-2, NNX09AG70G, NNX09AG63G, NNX09AJ79G, NNG05EC85C, Subcontract A991132BT, NNX09AP74A, NNX10AE46G, NNX09AW45G, and NNX14 AF43G, and NSF grant ATM-0904007. O.K. was supported by RFBR grant 14-02-00769 and partially by RFBR grant 14-02-00308. J.A.l.R. acknowledges support from NASA grants NNX14AF43G and NNX15AI65G. We are grateful to Valentina Zharkova for her interest in the work and valuable comments. O.K. thanks Bernard V. Jackson, Jack D. Scudder, and Leif Svalgaard for useful discussions.

REFERENCES

Ashour-Abdalla, M., El-Alaoui, M., Goldstein, M. L., et al. 2011, NatPh,7, 360
Bemporad, A. 2008, ApJ, 689, 572
Bian, N. H., & Kontar, E. P. 2013, PhRvL, 110, 151101
Bieber, J. W., Wanner, W., & Matthaeus, W. H. 1996, JGR, 101, 2511
Bothmer, V., & Schwenn, R. 1998, AnGeo, 16, 1
Cargill, P. J., Vlahos, L., Turkmani, R., Galsgaard, K., & Isliker, H. 2006, SSRv, 124, 249
Cartwright, M. L., & Moldwin, M. B. 2008, JGR, 113, A09105
Cartwright, M. L., & Moldwin, M. B. 2010, JGR, 115, A08102
Chian, A. C.-L., & Munoz, P. R. 2011, ApJL, 733, L34
Chollet, E. E., & Giacalone, J. 2008, ApJ, 688, 1368
Chollet, E. E., Mewaldt, R. A., Cummings, A. C., et al. 2010, JGR, 115, A12106
Crooker, N. U., Gosling, J. T., Bothmer, V., et al. 1999, SSRv, 89, 179
Crooker, N. U., Kahler, S. W., Larson, D. E., & Lin, R. P. 2004, JGR, 109, A03108
Czechowski, A., Strumik, M., Grygorczuk, J., et al. 2010, A&A, 516, A17
Desai, M. I., & Burgess, D. 2008, JGR, 113, A9
Drake, J. F., Swisdak, M., Che, H., & Shay, M. A. 2006a, Natur, 443, 553
Drake, J. F., Swisdak, M., Schoeffler, K. M., Rogers, B. N., & Kobayashi, S. 2006b, GeoRL, 33, 13105
Drake, J. F., Swisdak, M., & Fermo, R. 2013, ApJL, 763, L5
Eastwood, J. P., Balogh, A., Dunlop, M. W., & Smith, C. W. 2002, JGR,107, 1365
Enzl, J., Prech, L., Safrankova, J., & Nemecek, Z. 2014, ApJ, 796, 21
Eriksson, S., Newman, D. L., Lapenta, G., & Angelopoulos, V. 2014, PPCF, 56, 064008
Feng, H. Q., Wang, J. M., & Wu, D. J. 2012, ChSBu, 57, 1415
Fitzpatrick, R. 1999, NucFu, 33, 1049
Fitzpatrick, R., & Waelbroeck, F. L. 2012, PhPl, 19, 112501
Fitzpatrick, R., Waelbroeck, F. L., & Militello, F. 2006, PhPl, 13, 122507
Frank, A. G., & Satunin, S. N. 2014, JETPL, 100, 75
Goldstein, M. L., Matthaeus, W., & Ambrosiano, J. J. 1986, GeoRL, 13, 3205
Gómez-Herrero, R., Malandraki, O., Dresing, N., et al. 2011, JASTP, 73, 551
Gosling, J. T., Eriksson, S., Phan, T. D., et al. 2007, GeoRL, 34, L06102
Gosling, J. T., Skoug, R. M., Haggerty, D. K., & McComas, D. J. 2005a, GeoRL, 32, L14113
Gosling, J. T., Skoug, R. M., McComas, D. J., & Smith, C. W. 2005b, JGR, 110, A01107
Greco, A., Servidio, S., Matthaeus, W. H., & Dmitruk, P. 2010, P&SS,58, 1895
Guo, J. 2014, Ap&SS, 351, 159
Hoshino, M. 2012, PhRVL, 108, 135003
Jackson, B. V. 2012, Adv. in Geosciences: Solar and Terrestrial Science, Vol.20 (Singapore: World Scientific)
Janvier, M., Demoulin, P., & Dasso, S. 2014, SoPh, 289, 2633
Jensen, E. A., & Russell, C. T. 2009, GeoRL, 36, L05104
Kahler, S., & Lin, R. P. 1994, GeoRL, 21, 1575
Kahler, S., & Lin, R. P. 1995, SoPh, 161, 183
Khabarova, O., & Obridko, V. 2012, ApJ, 761, 82
Khabarova, O., & Zastenker, G. 2011, SoPh, 270, 311
Lanzerotti, L. J., & Sanderson, T. R. 2001, in The Heliosphere Near Solar Minimum: The Ulysses Perspective, ed. A. Balogh, R. G. Marsden & E. J. Smith (Chichester, UK: Praxis)
Lazarian, A., Eyink, G. L., Vishniac, E. T., & Kowal, G. 2015, Magnetic Fields in Diffuse Media, Vol. 407 (Berlin: Springer)
Lazarian, A., Vlahos, L., Kowal, G., et al. 2012, SSRv, 173, 557
Le, A., Egedal, J., Ng, J., et al. 2014, PhPl, 21, 012103
le Roux, J. A., Zank, G. P., Webb, G. M., & Khabarova, O. 2015, ApJ,801, 112
Li, G. 2008, ApJL, 672, L65

Li, G., Miao, B., Hu, Q., & Qin, G. 2011, PhRvL, 106, 125001
Li, G., Zank, G. P., & Rice, W. K. M. 2003, JGR, 108, 1082
Loureiro, N. F., Schekochihin, A. A., & Cowley, S. C. 2007, PhPl, 14, 100703
Malandraki, O. E., Lario, D., Lanzerotti, L. J., et al. 2005, JGR, 110, A09S06
Malandraki, O. E., Marsden, R. G., Tranquille, C., et al. 2008, JGR, 112, A06111
Manchester, W. B., IV, van der Holst, B., & Lavraud, B. 2014, PPCF, 56, 064006
Markidis, S., Henri, P., Lapenta, G., et al. 2013, PhPl, 20, 082105
Matthaeus, W. H., Ambrosiano, J. J., & Goldstein, M. L. 1984, PhRvL, 53, 1449
Matthaeus, W. H., Dasso, S., Weygand, J. M., et al. 2005, PhRvL, 95, 231101
Matthaeus, W. H., Goldstein, M. L., & Roberts, D. A. 1990, JGR, 95, 673
Merkin, V. G., Lyon, J. G., McGregor, S. L., & Pahud, D. M. 2011, GeoRL, 38, L14107
Moldwin, M. B., Ford, S., Lepping, R., Slavin, J., & Szabo, A. 2000, GeoRL, 27, 57
Moldwin, M. B., & Hughes, W. J. 1992, JGR, 97, A12
Moldwin, M. B., Phillips, J. L., Gosling, J. T., et al. 1995, JGR, 100, 19903
Murphy, N., Smith, E. J., Burton, M. E., Winterhalter, D., & McComas, D. J. 1993, JPL Tech. Rep., 20060039442
Mursula, K., & Hiltula, T. 2003, GeoRL, 30, 2135
Oka, M., Phan, T.-D., Krucker, S., Fujimoto, M., & Shinohara, I. 2010, ApJ, 714, 915
Osman, K. T., Matthaeus, W. H., Gosling, J. T., et al. 2014, PhRvL, 112, 215002
Pritchett, P. L. 2006, GeoRL, 33, 13104
Pritchett, P. L. 2007, PhPl, 14, 052102
Pritchett, P. L. 2008, PhPl, 15, 102105
Reames, D. V. 2009, ApJ, 693, 812
Reames, D. V. 2013, SSRv, 175, 53
Richardson, I. G. 2004, SSRv, 111, 267
Richardson, J. D., Stone, E. C., Cummings, A. C., et al. 2006, GeoRL, 33, L21112
Sanderson, T. R. 1997, Solar and Heliospheric Plasma Physics, Vol. 489 (Berlin: Springer)
Scudder, J. D., Holdaway, R. D., Daughton, W. S., et al. 2012, PhRvL, 108, 225005
Sitnov, M. I., Merkin, V. G., Swisdak, M., et al. 2014, JGR, 119, 7151
Shibata, K., & Tanuma, S. 2001, EP&S, 53, 473
Smith, C. W., Matthaeus, W. H., Zank, G. P., et al. 2001, JGR, 106, 8253
Sonnerup, B. U. O., & Cahill, L. J., Jr. 1968, JGR, 73, 1757
Susino, R., Bemporad, A., & Krucker, S. 2013, ApJ, 777, 93
Tessein, J. A., Matthaeus, W. H., Wan, M., et al. 2013, ApJL, 776, L8
Tokumaru, M. 2013, PJAB, 89, 2
Verkhoglyadova, O. P., Li, G., Ao, X., & Zank, G. P. 2012, ApJ, 757, 75
Verkhoglyadova, O. P., Li, G., Zank, G. P., et al. 2010, JGR, 115, A12103
Waelbroeck, F. L. 2009, NucFu, 49, 104025
Waelbroeck, F. L., & Fitzpatrick, R. 1997, PhRvL, 78, 1703
Wei, F., Hu, Q., Feng, X., & Schwen, R. 2000, ScChA, 43, 629
Xu, X. Q., Nevins, W. M., Cohen, R. H., Myra, J. R., & Snyder, P. B. 2002, NJPh, 4, 53
Zank, G. P., le Roux, J. A., Webb, G. M., Dosch, A., & Khabarova, O. 2014, ApJ, 797, 28
Zank, G. P., Li, G., Florinski, V., et al. 2004, JGR, 109, A04107
Zank, G. P., Li, G., & Verkhoglyadova, O. 2007, SSRv, 130, 255
Zank, G. P., & Matthaeus, W. H. 1992, JGR, 97, 17189
Zank, G. P., & Matthaeus, W. H. 1993, PhFl, A5, 257
Zank, G. P., Rice, W. K. M., & Wu, C. C. 2000, JGR, 105, 25079
Zelenyi, L. M., Lominadze, J. G., & Taktakishvili, A. L. 1990, JGR, 95, 3883
Zharkova, V. V., Arzner, K., Benz, A. O., et al. 2011, SSRv, 159, 357
Zharkova, V. V., & Khabarova, O. V. 2012, ApJ, 752, 35
Zharkova, V. V., & Khabarova, O. V. 2015, AnGeo, 33, 457